\documentclass[11pt,a4paper,twoside]{article}
\pdfoutput=1

\usepackage[a4paper,left=0.9in,right=0.9in,top=0.85in,bottom=1in,marginparwidth=0pt]{geometry}
\usepackage{newpxtext}
\usepackage[vvarbb]{newpxmath}
\usepackage{amsmath,amsthm,mathtools}
\usepackage{thm-restate}
\usepackage{xspace}

\usepackage[T1]{fontenc}
\usepackage[utf8]{inputenc}

\usepackage[tracking=true,protrusion=true,expansion=true]{microtype}
\usepackage{enumitem}
\setlist{itemsep=2pt,topsep=5pt,parsep=0pt}
\usepackage{comment}

\usepackage{xcolor}
\definecolor{accent}{HTML}{1D4E89}
\definecolor{accentmuted}{HTML}{5B84B1}
\definecolor{ink}{HTML}{1A1A1A}
\definecolor{muted}{HTML}{5A5A5A}
\color{ink}

\colorlet{figA}{accent}
\definecolor{figB}{HTML}{C8551A}
\definecolor{figC}{HTML}{6B4C9A}
\definecolor{figD}{HTML}{3E8E6A}

\newtheoremstyle{modernplain}{0.9\baselineskip}{0.9\baselineskip}{\itshape}{0pt}{\bfseries\color{accent}}{.}{0.6em}{}
\newtheoremstyle{moderndef}{0.9\baselineskip}{0.9\baselineskip}{\normalfont}{0pt}{\bfseries\color{accent}}{.}{0.6em}{}
\theoremstyle{modernplain}
\usepackage{thm-restate}
\usepackage{thmtools}

\makeatletter
\newskip\thmt@rst@savedskip
\let\thmt@rst@storecounters@orig\thmt@rst@storecounters
\def\thmt@rst@storecounters#1{%
  \thmt@rst@savedskip\lastskip
  \thmt@rst@storecounters@orig{#1}%
  \ifvmode\ifdim\thmt@rst@savedskip=\z@\else
    \nobreak\vskip-\thmt@rst@savedskip\vskip\thmt@rst@savedskip
  \fi\fi
}
\makeatother

\usepackage{titlesec}
\titleformat{\section}{\normalfont\Large\bfseries\color{accent}}{\thesection}{0.6em}{}
\titleformat{\subsection}{\normalfont\large\bfseries\color{accent}}{\thesubsection}{0.6em}{}
\titleformat{\paragraph}[runin]{\normalfont\bfseries\color{accent}}{}{0pt}{}[\ ]
\titlespacing*{\section}{0pt}{1.8em}{0.8em}
\titlespacing*{\subsection}{0pt}{1.4em}{0.6em}
\titlespacing*{\paragraph}{0pt}{1.4em}{0.8em}

\usepackage[labelfont={bf,color=accent},font=small,labelsep=period]{caption}

\usepackage{fancyhdr}
\fancypagestyle{plain}{%
  \fancyhf{}
  \fancyfoot[C]{\footnotesize\color{muted}\thepage}
  
}

\makeatletter
\renewenvironment{proof}[1][\proofname]{\par\pushQED{\qed}%
  \normalfont\topsep6\p@\@plus6\p@\relax
  \trivlist
  \item[\hskip\labelsep\color{accent}\bfseries #1\@addpunct{.}]\ignorespaces
}{\popQED\endtrivlist}
\makeatother

\usepackage{xspace}
\usepackage{xparse}
\usepackage{algorithm}
\usepackage{mathtools}
\usepackage[commentColor=black]{algpseudocodex}

\makeatletter
\newcommand{\@authorlist}{}
\newcounter{authorcount}
\RenewDocumentCommand{\author}{m g g}{%
  \ifnum\value{authorcount}>0 \g@addto@macro\@authorlist{\hspace{2.5em}}\fi
  \stepcounter{authorcount}%
  \g@addto@macro\@authorlist{%
    \begin{minipage}[t]{0.42\textwidth}\centering
      \textbf{#1}%
      \IfNoValueTF{#2}{}{\\[3pt]{\small\itshape #2}}%
      \IfNoValueTF{#3}{}{\\[2pt]{\small\href{mailto:#3}{\nolinkurl{#3}}}}%
    \end{minipage}%
  }%
}
\newcommand{\@fundingextra}{}
\newcommand{\keywords}[1]{\renewcommand{\@keywordstext}{#1}}
\newcommand{\@keywordstext}{}
\newcommand{\funding}[1]{\renewcommand{\@fundingtext}{#1}}
\newcommand{\@fundingtext}{}

\renewcommand{\maketitle}{%
  \begingroup
  \thispagestyle{plain}
  \begin{center}
    {\fontsize{19}{23}\selectfont\bfseries\color{accent}\@title\par}
    \vspace{1.3em}
    {\setlength{\lineskip}{1.6em}\@authorlist\par}
  \end{center}
  \vspace{1.3em}
  {\color{accent}\hrule height0.6pt}
  \endgroup
}

\renewenvironment{abstract}{%
  \vspace{1.2em}
  \noindent{\bfseries\scshape\color{accent}\small Abstract}\par\vspace{0.5em}
  \small\noindent\ignorespaces
}{%
  \par\vspace{1.2em}
  \small
  \ifx\@keywordstext\@empty\else\noindent{\bfseries\color{accent}Keywords.}~\@keywordstext\par\vspace{4pt}\fi
  \ifx\@fundingtext\@empty\else\noindent{\bfseries\color{accent}Funding.}~\@fundingtext\@fundingextra\par\vspace{4pt}\fi
  \vspace{0.6em}
  {\color{accent}\hrule height0.6pt}
  \vspace{1.6em}
}
\makeatother

\usepackage[colorlinks,linkcolor=accent,citecolor=accent,urlcolor=accent,anchorcolor=accent,filecolor=accent]{hyperref}

\usepackage{hyperref}
\usepackage[nameinlink,capitalise]{cleveref}
\crefname{claim}{Claim}{Claims}

\crefname{ALG@line}{line}{lines}
\makeatletter
\g@addto@macro\ALG@step{\gdef\@currentcounter{ALG@line}}
\AtEndEnvironment{algorithmic}{\gdef\@currentcounter{algorithm}}
\makeatother

\newtheorem{theorem}{Theorem}
\newtheorem{lemma}[theorem]{Lemma}
\newtheorem{claim}[theorem]{Claim}
\theoremstyle{moderndef}
\newtheorem{definition}[theorem]{Definition}

\hypersetup{bookmarksdepth=subsection}

\DeclareMathOperator{\E}{\mathbf{E}}
\newcommand{\C}{\mathcal{C}}
\DeclarePairedDelimiter{\mult}{\langle}{\rangle_q}
\DeclarePairedDelimiter{\genmult}{\langle}{\rangle}
\DeclarePairedDelimiter{\set}{\{}{\}}

\renewcommand{\P}{\mathcal{P}}
\newcommand*{\Nat}{\mathbb{N}}
\newcommand*{\Int}{\mathbb{Z}}
\newcommand{\OPT}{\textsc{Opt}\xspace}
\newcommand{\ALG}{\textsc{Alg}\xspace}
\newcommand{\ICB}{\textsc{ICB}\xspace}
\newcommand{\ICBss}{\ensuremath{\textsc{ICB}^\textnormal{ss}}\xspace}
\newcommand{\ICBrb}{\ensuremath{\textsc{ICB}^\textnormal{rb}}\xspace}
\newcommand{\dist}{\textrm{dist}}
\newcommand{\size}{\textrm{size}}
\newcommand{\cnt}{\textrm{cnt}}
\newcommand{\F}{\mathcal{F}}
\newcommand{\I}{\mathcal{I}}
\newcommand{\suchthat}{\ensuremath{\,:\,}}
\newcommand{\Epoch}{\mathcal{E}}
\newcommand{\eps}{\varepsilon}
\newcommand{\define}{\coloneqq}

\begin{document}

\title{An Improved Subquadratic Bound for Online Bisection}

\author{Marcin Bienkowski}{Institute of Computer Science, University of Wroc\l{}aw, Poland}{marcin.bienkowski@cs.uni.wroc.pl}

\author{Stefan Schmid}{TU Berlin and Weizenbaum Institute, Berlin, Germany}{stefan.schmid@tu-berlin.de}

\maketitle

\begingroup
\renewcommand{\thefootnote}{}
\footnotetext{A preliminary version of this paper appeared in the proceedings of the 41st International Symposium on Theoretical Aspects of Computer Science (STACS 2024).}
\endgroup

\begin{abstract}
The online bisection problem is a natural dynamic variant of the classic bisection problem, where one has to dynamically maintain a partition of $n$ elements into two clusters of cardinality $n/2$. During runtime, an~online algorithm is given a sequence of requests, each being a pair of elements: an~inter-cluster request costs one unit while an intra-cluster one is free. The algorithm may change the partition, paying a unit cost for each element that changes its cluster. This natural problem admits a simple deterministic $O(n^2)$-competitive algorithm [Avin et al., DISC 2016]. While several significant improvements over this result have been obtained since the original work, all of them either limit the generality of the input or assume some form of resource augmentation (e.g., larger clusters). Moreover, the algorithm of Avin et al.\ achieves the best known competitive ratio even if randomization is allowed. In this paper, we present the first randomized online algorithm that breaks this natural quadratic barrier and achieves a competitive ratio of $\tilde{O}(n^{7/4})$ without resource augmentation and for an arbitrary sequence of requests.
\funding{This work was supported by the Polish National Science Centre grant 2022/45/B/ST6/00559 and by the European Research Council (ERC) Proof-of-Concept grant 101287293 (FortifyNet).}
\end{abstract}


\section{Introduction}

The clustering of elements into subsets that are related by some similarity measure is a~fundamental algorithmic problem. The problem arises in multiple contexts: a~well-known abstraction is \emph{the bisection problem}~\cite{Krauth16} that asks for a partitioning of $n$ graph nodes (elements) into two clusters of size $n/2$, so that the number of graph edges in the cut is minimized. This problem is NP-hard and its approximation ratio has been improved in a~long line of papers~\cite{SarVaz95,ArKaKa99,FeKrNi00,FeiKra02,KraFei06}; the currently best approximation ratio of $O(\log n)$ was given by R{\"a}cke~\cite{Raecke08}. 

Recently this problem has been studied in a \emph{dynamic variant}~\cite{AvLoPS16,RaScZa22}, where instead of a~fixed graph, we are given a sequence of element pairs. Serving a pair of elements that are in different clusters costs one unit, while a request between two elements in the same cluster is free. After serving a request, an algorithm may modify the partition, paying a unit cost for each element that changes its cluster. 

A natural motivation for this problem originates from data centers where communicating virtual machines (elements) have to be partitioned between servers (clusters) and the overall communication cost has to be minimized: by collocating virtual machines on the same server, their communication becomes free, while the communication between virtual machines in different clusters involves using network bandwidth. Modern virtualization technology supports the seamless migration of virtual machines between servers, but migrations still come at the cost of data transmission. The goal is to minimize data transmission (across the network), comprising inter-cluster communication and migration.

This practical application motivates yet another aspect of the problem: the communication pattern (and, in particular, the sequence of communication pairs) is typically not known ahead of time. Accordingly, we study the dynamic variant in the \emph{online setting}, where the sequence of communication pairs is not known a priori to an online algorithm \ALG: \ALG has to react immediately without the knowledge of future communication requests. To evaluate its performance, we use a~standard notion of \emph{competitive ratio}~\cite{BorElY98} that compares the cost of \ALG with the cost of the optimal (offline) solution \OPT: the \ALG-to-\OPT cost ratio is subject to minimization.


\subsection{Previous results}

Avin et al.~introduced the online bisection problem and presented a~simple deterministic online algorithm that achieves the~competitive ratio of $O(n^2)$~\cite{AvLoPS16}. Their algorithm belongs to a~class of component-preserving algorithms (formally defined in \cref{sec:preliminaries}). Roughly speaking, it splits the request sequence into epochs. Within a~single epoch, it glues requested element pairs together creating components and assigns all elements of any component to the same cluster. If such an~assignment is no longer feasible, i.e., components cannot be preserved (kept on the same cluster), an epoch terminates. It is easy to argue (cf.~\cref{sec:preliminaries}) that \OPT pays at least $1$ in an epoch, and any component-preserving algorithm pays at most~$n^2$, thus being $n^2$-competitive. 

Perhaps surprisingly, no better algorithm (even a randomized one) has been known for the online bisection problem. On the negative side, a lower bound of $\Omega(n)$~\cite{AvBLPS20} for deterministic algorithms follows by the reduction from online paging~\cite{SleTar85}, and the best known lower bound for randomized algorithms is merely $\Omega(\log n)$~\cite{HeNeRS21} (it follows by repeating their construction for the so-called learning variant multiple times).


\subsection{Our contribution}

We present the first algorithm for the online bisection problem that beats the quadratic competitive ratio. All previous results with better ratios required some relaxation: either used resource augmentation or restricted the generality of the input sequence. Our \textsc{Improved Component Based} algorithm (\ICB) is randomized, follows the component-preserving framework outlined above, and achieves the competitive ratio of $O(n^{7/4} \cdot \log^{1/4} n)$.


\subsection{Our algorithmic ideas}

Assume that an algorithm follows the component-preserving framework and we want to improve its cost within a single epoch. We may look at the problem more abstractly: there is a set of ``allowed'' partitions (the ones that map elements to clusters in a~component-preserving way), and this set is constantly shrinking. Consider an~algorithm that, whenever it needs to change its partition, changes it to one chosen uniformly at random from the set of still allowed partitions. Using standard arguments, we may show that the algorithm changes its partition at most $O(\log y)$ times within an epoch, where $y$ is the number of ``allowed'' partitions at the beginning. As the cost of serving a request and the cost of changing the partition sum up to at most $1+n$, the overall cost of such a routine is $O(n \cdot \log y)$. 

At the beginning of an epoch, $y = \binom{n}{n/2}$, and thus $O(n \cdot \log y) = O(n^2)$. That is, the randomized routine itself would fail to beat the quadratic upper bound of~\cite{AvLoPS16} if it is applied to the entire epoch. However, we may execute it in the second stage of an epoch, once the number of ``allowed'' partitions drops appropriately.

In the first stage of an epoch, our proposed algorithm \ICB carefully tracks the component sizes. In a single step, it needs to merge two components into a single one and to map all elements of the resulting component to the same cluster. To this end, it usually has to move one of the merged components to the other cluster. A crucial insight is that, most of the time, the size of the moved component can be expressed as a~\emph{non-negative} integer combination, with small coefficients, of a~moderate number of existing component sizes: in such a case, only a~limited number of existing components have to change their clusters, and the cardinalities of both clusters remain unchanged.

Converting this intuition into an actual algorithm is not easy. The main difficulty is that the set of component sizes used in such combinations changes over time, and \ICB has to keep track of these changes. To this end, \ICB maintains a~\emph{GCD estimator}: the greatest common divisor of those component sizes that are \emph{small} (below a threshold~$q$) and sufficiently well-represented among the current components. We show that this estimator, unlike the true GCD, changes only $O(\log q)$ times within an epoch. We then use number-theoretic properties to argue that whenever \ICB merges two components into a single one, then usually one of them has a size divisible by the current value of the estimator, and thus the resulting repartitioning incurs the movement of only a~moderate number of other components. 

The low-cost argument depends, however, on the property that not only are there many components of sizes divisible by the GCD, but also both clusters contain sufficiently many of them: as the clusters have to remain of equal cardinality, components must be moved in both directions. \ICB ensures this property by regularly running a~``rebalancing'' routine. At some point, maintaining this property is no longer possible. We prove that such failure guarantees that the total number of ``allowed'' partitions is appropriately low: \ICB switches then to the second stage of an epoch, where it executes the randomized policy outlined above. 

We emphasize that the last implication is crucial for our approach. The failure of the rebalancing routine guarantees that the randomized routine of the second stage is already cheap. In particular, \ICB does not have to guess when to switch stages: it runs the first stage as long as rebalancing is possible, and it switches to the second stage in the first step when it is not.


\subsection{Comparison with the conference version}

A preliminary version of this paper appeared at STACS 2024~\cite{BieSch24}, where we showed the competitive ratio of $O(n^{23/12} \cdot \log^{1/2} n)$. 

In this paper, we improve three parts of the analysis. We sharpen the number-theoretic bound of \cref{cla:number_theory}, which yields smaller coefficients in the linear combinations, and thus fewer elements moved in a~single step of the first stage. Moreover, in the conference version, a component size was sufficiently well represented if the number of components of this size was at least a fixed threshold~$w$. Here we replace $w$ by a~function $w(i)$ that depends on the component size~$i$, i.e., we require fewer components of larger sizes. Finally, we bound the number of rebalancing events more carefully: two consecutive rebalancing events are separated either by a step of a rare type or by sufficiently many regular steps.

These three changes allow us to choose better values of the algorithm parameters, which leads to the improved bound of $O(n^{7/4} \cdot \log^{1/4} n)$.


\subsection{Related work}

The lack of progress toward improving the $O(n^2)$ upper bound motivated the investigation of simplified variants. 

A natural relaxation involves resource augmentation, where each cluster of an~online algorithm can accommodate $(1+\eps) \cdot (n/2)$ elements. The performance of an~online algorithm is compared to \OPT whose clusters both still have capacity~$n/2$. Surprisingly, the competitive ratio remains~$\Omega(n)$ even for large $\eps$ (but as long as \mbox{$\eps < 1$})~\cite{AvBLPS20}. On the positive side, Rajaraman and Wasim showed an~$O(n \log n)$-competitive deterministic algorithm for a~fixed~$\eps > 0$~\cite{RajWas22}.

Another relaxation was introduced by Henzinger et al.~\cite{HeNeSc19} who initiated the study of the so-called \emph{learning variant}. In this variant, there exists a fixed partition $\bar{p}$ (unknown to an~algorithm), and all requests are consistent with $\bar{p}$ (i.e., given between same-cluster pairs). Clearly, the optimal solution simply changes its partition to $\bar{p}$ at the very beginning. The deterministic variant is asymptotically resolved: the optimal competitive ratio is~$\Theta(n)$~\cite{PaPaSc20,PaPaSc21}. For the model where the learning variant is combined with resource augmentation, Henzinger et al.~determined the optimal competitive ratio (for any fixed $\eps > 0$)~\cite{HeNeRS21}: for two clusters, it is $\Theta(\log n)$, which is attained already by a~deterministic algorithm, and the matching lower bound holds even for randomized ones. A more general learning model where the offline algorithm is not restricted to keep connected components together has recently been considered~\cite{RaScZa25}.

The online bisection problem has also been studied in a generalized form, where there are $\ell > 2$ clusters, each of size $n / \ell$. This extension is usually referred to as \emph{online balanced graph partitioning}. Some of the results presented above can be generalized to this
variant~\cite{AvBLPS20,AvLoPS16,HeNeRS21,HeNeSc19,PaPaSc20,PaPaSc21,RajWas22}. This generalization was investigated also in models with a large augmentation of $\eps > 1$~\cite{AvBLPS20,FoRaSc21,HeNeRS21,RaScZa22} and in settings with small (or even constant-size) clusters~\cite{AvBLPS20,BBKRSV21,PaPaSc21}. Finally, improved bounds were given for restricted input sequences, e.g., where every request must be an edge of a fixed cycle spanning all nodes~\cite{BaBiET26,RaScZa23}.


\section{Preliminaries}
\label{sec:preliminaries}

We have a set $V$ of $n$ elements (we assume that $n$ is even) and two clusters $0$ and~$1$. A~\emph{partition} of these elements is a mapping $p: V \to \{0,1\}$ such that $|p^{-1}(0)| = |p^{-1}(1)| = n/2$, i.e., each cluster contains exactly $n/2$ elements. For two partitions $p$ and~$p'$, we use $\dist(p,p') = |\{ v \in V \suchthat p(v) \neq p'(v) \}|$ to denote the number of elements that change their clusters when switching from partition $p$ to~$p'$.


\subsection{Problem definition}

An input for the online bisection problem consists of an initial partition~$p_0$ and a~sequence of element pairs $((u_t, v_t))_{t \geq 1}$. In step $t \geq 1$, an online algorithm \ALG is given a~pair of elements $(u_t, v_t)$: it pays a~\emph{service cost} of $1$ if $p_{t-1}(u_t) \neq p_{t-1}(v_t)$ and~$0$~otherwise. Afterward, \ALG has to compute a new partition $p_t$ (possibly $p_t = p_{t-1}$) and pay $\dist(p_{t-1}, p_t)$ for changing partition $p_{t-1}$ to partition~$p_t$. 

For an input $\I$ and an online algorithm $\ALG$, we use $\ALG(\I)$ to denote its total cost on~$\I$, whereas $\OPT(\I)$ denotes the optimal cost of an offline solution. $\ALG$ is \emph{$\gamma$-competitive} if there exists~$\beta$, such that $\ALG(\I) \leq \gamma \cdot \OPT(\I) + \beta$ for any input $\I$. While $\beta$~has to be independent of $\I$, it may be a~function of $n$. For a randomized algorithm \ALG, we replace $\ALG(\I)$ with its expectation $\E[\ALG(\I)]$, taken over all random choices of \ALG.


\subsection{Component-preserving framework}

A~natural way of tackling the problem is to split requests into epochs. In a single epoch, an~online algorithm \ALG treats requests as edges connecting requested element pairs. Edges in a~single epoch induce connected components of elements. A \emph{component-preserving} algorithm always keeps elements of each component in the same cluster. If it is no longer possible, the current epoch ends, all edges are removed (each element is now in its own singleton component), and a new epoch begins with the next step. We note that the currently best $O(n^2)$-competitive deterministic algorithm of~\cite{AvLoPS16} is component-preserving. 

Now, we recast the online bisection problem assuming that we analyze a~component-preserving algorithm \ALG. First, observe that \ALG completely ignores all intra-component requests (they also incur no cost on \ALG). Consequently, we may assume that the input for \ALG (within a single epoch) is a sequence of component sets $\C_t$, where:
\begin{itemize}
\item $\C_0$ is the initial set of $n$ singleton components;
\item $\C_t$ (presented in step $t \geq 1$) is created from $\C_{t-1}$ by merging two of its components, denoted $c^t_x$ and $c^t_y$. They are merged into a component, denoted $c^t_z$, i.e.,
    \[
        \C_t = \C_{t-1} \cup \, \set{c^t_z} \setminus \set{c^t_x,c^t_y}.
    \]
\end{itemize}

For a given set of components $\C$, let $\P(\C)$ denote the set of all \emph{$\C$-preserving partitions} of $n$~elements into two clusters, i.e., ones that place all elements of a single component of $\C$ in the same cluster. In response to $\C_t$, \ALG chooses a $\C_t$-preserving partition $p_t$. If $\P(\C_t)$ is empty though, then \ALG does not change its partition, and an epoch terminates. The following observations let us focus on $\ALG$'s behavior in a~single epoch only.

\begin{lemma}
\label{lem:partition_preserving_trivial}
    The epoch of any component-preserving algorithm contains at most $n-1$ steps, and the single-step cost is at most $n+1$.
\end{lemma}

\begin{proof}
    In each step, the number of components decreases. Thus, after $n-1$ steps, all elements would be in the same component, and hence $\P(\C_{n-1}) = \emptyset$. In a single step, an~algorithm pays at most $1$ for serving the request and changes the cluster of at most $n$ elements.
\end{proof}

\begin{lemma}
\label{lem:epoch_cost_only}
    If a component-preserving algorithm \ALG pays at most $R$ in expectation in any epoch, then \ALG is $R$-competitive.
\end{lemma}

\begin{proof}
    Fix any finished epoch $\Epoch$ in an input (any epoch except possibly the last one). The final step of $\Epoch$ serves as a certificate that any algorithm keeping a static partition throughout~$\Epoch$ has a~non-zero cost. On the other hand, changing partition costs at least $1$, and thus \mbox{$\OPT(\Epoch) \geq 1$}. 

    The lemma follows by summing the expected costs over all epochs except the last one. Observe that the cost of the last epoch is at most $n^2-1$ (by \cref{lem:partition_preserving_trivial}), and thus can be placed in the additive term $\beta$ in the definition of the competitive ratio (cf.~\cref{sec:preliminaries}).
\end{proof}

Note that by \cref{lem:partition_preserving_trivial} and \cref{lem:epoch_cost_only} the competitive ratio of any component-preserving algorithm (including that of~\cite{AvLoPS16}) is at most $(n-1) \cdot (n+1) < n^2$.


\subsection{Notation}

For an integer $\ell$, we define $[\ell] = \{1, 2, \ldots, \ell\}$. For any finite set $A \subset \Nat_{> 0}$, we use $\gcd(A)$ to denote the greatest common divisor of all integers from $A$; we assume that $\gcd(\emptyset) = \infty$. All logarithms are binary.


\paragraph{Component sizes.}

For any component~$c$, we denote its size (number of elements) by $\size(c)$. Fix any component set $\C$ and an integer $i$. Let $\cnt_i(\C) \define | \{ c \in \C \suchthat \size(c) = i \} |$ denote the number of components in $\C$ of size $i$.

Next, fix a partition $p \in \P(\C)$ and a cluster $y \in \{0,1\}$. As $p$ is $\C$-preserving, it is constant on all elements of a given component $c \in \C$, and thus we may extend $p$ to components from~$\C$. We define 
\[ 
    \cnt_i(\C,p,y) \define | \{ c \in \C \suchthat p(c) = y \,\wedge\, \size(c) = i \} |
\] 
as the number of components in $\C$ of size $i$ that are inside cluster $y$ in partition $p$.

We extend both notions to sets of sizes, i.e., for any set $A$, we set $\cnt_A(\C) \define \sum_{i \in A} \cnt_i(\C)$ and $\cnt_A(\C,p,y) \define \sum_{i \in A} \cnt_i(\C,p,y)$.


\paragraph{Balanced partitions.}

Fix a set of components $\C$, a~set $A$, and an integer $\ell$. For a given partition $p \in \P(\C)$, we say that $p$ is $(A,\ell)$-balanced if 
\begin{align*}
    \textstyle \cnt_A(\C,p,y) \geq \ell \qquad \text{for each $y \in \{0,1\}$}. 
\end{align*}
That is, an $(A,\ell)$-balanced partition $p$ of $\C$ keeps at least $\ell$ components of sizes in $A$ in each cluster. We use $\P(\C,A,\ell) \subseteq \P(\C)$ to denote the set of all $(A,\ell)$-balanced $\C$-preserving partitions.


\paragraph{Algorithm parameters.}

Our algorithm \ICB uses two integer parameters described below whose precise values will be chosen later. 
\begin{itemize}
    \item Parameter $q \in [n]$. A component size is \emph{large} if it is greater than $q$ and is \emph{small} otherwise.
    \item Parameter $d \in [n]$. It is the number of components of sizes from the current core set (defined later) that \ICB tries to keep in each cluster.
\end{itemize}
Furthermore, \ICB uses the function $w: [n] \to \mathbb{R}_{\geq 0}$ where
\[
    w(i) = 4 \cdot q^2 / i + 1.
\]
$\ICB$ works with any values of $q$, $d$, as long as they satisfy 
\begin{align}
    \label{eq:parameter_values_relations}
    d \geq \sum_{i \in [q]} (2 \cdot w(i) + 3) && \text{and} && 2 \cdot d \leq n.
\end{align} 
The parameter values yielding the competitive ratio of $O(n^{7/4} \cdot \log^{1/4} n)$ are chosen in \cref{thm:competitive_ratio}.


\paragraph{Divisibility.}

For succinctness of arguments, we extend the notion of divisibility. Recall that $a \mid b$ means that $b$ is divisible by~$a$ and is well defined for any two positive integers $a$ and $b$. We extend it also to the cases where $a$ and $b$ are possibly infinite: $a \mid \infty$ for any $a \in \Nat_{> 0} \cup \{\infty\}$ and $\infty \nmid b$ for any $b \in \Nat_{>0}$.

\begin{claim}
\label{lem:gcd_property}
    For any sets of integers $A$ and $B$, it holds that $\gcd(A) \mid \gcd(A \cap B)$.
\end{claim}

\begin{proof}
    The claim follows trivially if $A \cap B = \emptyset$ as in such case $\gcd(A \cap B) = \infty$. Thus, we may assume that $A \cap B \neq \emptyset$ (and hence also $A \neq \emptyset$). Fix any $i \in A \cap B$: as $i \in A$, we have $\gcd(A) \mid i$. Hence, $\gcd(A)$ is a divisor of all numbers from $A \cap B$, and therefore $\gcd(A) \mid \gcd(A \cap B)$.
\end{proof}


\paragraph{Multiples.}

For any $k \in \Nat_{>0} \cup \{\infty\}$, we define 
\[ 
    \mult{k} \define \{ \ell \cdot k \suchthat \ell \in \Nat \} \cap [q].
\]
In particular, $\mult{\infty} = \emptyset$. Observe that 
\begin{align}
\label{eq:mult_gcd}
    k = \gcd(\mult{k}) \qquad \text{for each $k \in [q] \cup \{\infty\}$}.
\end{align}


\section{Cheap merges}

In the following we prove that in certain cases, it is possible to merge two components of a given component set $\C$ into a single one, while changing the clusters of only a~moderate number of elements.

We start with the following number-theoretic claim, proven in \cref{sec:proofs}. 

\begin{claim}
\label{cla:number_theory}
    Fix two non-empty disjoint sets of positive integers $A = \set{a_1,\dots,a_k}$ and $B = \set{b_1,\dots,b_\ell}$, and a non-negative integer $x$
    such that $\gcd(A \uplus B) \mid x$.
    Then, there exist non-negative integers $\alpha_1,\dots,\alpha_k$ and $\beta_1,\dots,\beta_\ell$, such that
    \[
        \sum_{i=1}^k \alpha_i \cdot a_i
        =
        x + \sum_{j=1}^\ell \beta_j \cdot b_j.
    \]
    Moreover, both sides of the above equation are at most $4 \cdot H^2$, 
    where $H = \max (A \cup B \cup \set{x})$.
\end{claim}


\begin{claim}
\label{cla:balanced_partition}
    Fix a component set $\C$ and let $p$ be a $\C$-preserving partition. Fix $A \subseteq [n]$ with $|A| \geq 2$, and let $z : [n] \to \mathbb{R}_{\geq 0}$ be a real-valued function. If
    \begin{itemize}
        \item $\cnt_i(\C) \geq 2 \cdot z(i)$ for each $i \in A$, and
        \item $\cnt_A(\C, p, y) \geq \sum_{i \in A} z(i)$ for each cluster $y \in \{0, 1\}$,
    \end{itemize}
    then $A$ can be partitioned into two disjoint and non-empty sets $A_0, A_1$, such that $\cnt_i(\C, p, y) \geq z(i)$ for each $y \in \{0, 1\}$ and for each $i \in A_y$.
\end{claim}

\begin{proof}
    For each cluster $y \in \set{0, 1}$, let 
    \begin{align*}
        A'_y & \define \set{ i \in A \suchthat \cnt_i(\C, p, y) \geq z(i)}.
    \end{align*}
    Fix $i \in A$. As $\cnt_i(\C) = \cnt_i(\C,p,0) + \cnt_i(\C,p,1) \geq 2 \cdot z(i)$, we have $\cnt_i(\C,p,y) \geq z(i)$ for at least one $y \in \set{0,1}$. That is, component size $i$ belongs to at least one of the sets $A'_0$ or~$A'_1$, and hence $A'_0 \cup A'_1 = A$. 
    
    Now fix $y \in \set{0,1}$. By the claim assumption, $\sum_{i \in A} \cnt_i(\C, p, y) \geq \sum_{i \in A} z(i)$, and thus there exists $i \in A$ such that $\cnt_i(\C, p, y) \geq z(i)$. This implies $A'_y \neq \emptyset$.
    
    If $A'_0$ and $A'_1$ are disjoint, then $A_0 = A'_0$ and $A_1 = A'_1$ satisfy all the required conditions. 
    
    Otherwise, let $j$ be an arbitrary element of $A'_0 \cap A'_1$. We consider four disjoint sets: $B_0 = A'_0 \setminus A'_1$, $B_1 = A'_1 \setminus A'_0$, $B_2 = \set{j}$, and $B_3 = A'_0 \cap A'_1 \setminus \set{j}$. As the two clusters play symmetric roles, we may assume without loss of generality that $|B_0| \leq |B_1|$. We set $A_0 = B_0 \uplus B_2$ and $A_1 = B_1 \uplus B_3$. Note that $A_0 \subseteq A'_0$ and $A_1 \subseteq A'_1$, which yields the required lower bounds on $\cnt_i(\C,p,y)$. Clearly, $A_0$ and $A_1$ are disjoint, $A = A_0 \uplus A_1$, and $|A_0| \geq 1$, i.e., $A_0$ is non-empty. To show that $A_1$ is non-empty as well, we consider two cases. If $|A_0| = 1$, then we have $|A_1| = |A| - |A_0| \geq 2 - 1 = 1$. Otherwise $|A_0| \geq 2$, but then $|A_1| \geq |B_1| \geq |B_0| = |A_0| - 1 \geq 1$. 
\end{proof}

\begin{lemma}
\label{lem:regular_merge}
    Fix a non-empty set $A \subseteq [q]$ and two consecutive component sets $\C_{t-1}$ and~$\C_t$. Suppose that 
    \begin{itemize}
        \item $\cnt_i(\C_t) \geq 2 \cdot w(i) + 1$ for each $i \in A$,
        \item $\size(c^t_x) \in \mult{\gcd(A)}$ or $\size(c^t_y) \in \mult{\gcd(A)}$, and
        \item partition $p_{t-1}$ is $\C_{t-1}$-preserving and $(A, \sum_{i \in A} w(i))$-balanced.
    \end{itemize}
    Then there exists a $\C_t$-preserving partition $p$, such that $\dist(p_{t-1},p) = O(q^2)$.
\end{lemma}

The intuition behind this lemma is that if one of the merged components, say~$c^t_x$, has a~size that is divisible by $\gcd(A)$, then its size can be expressed as a linear combination of other component sizes in $A$. This means that when we move $c^t_x$ to the other cluster and change the clusters of the components dictated by the linear combination, the total number of elements on both sides remains unchanged. The crux of the proof is to show that (i) the number of moved components is small (the coefficients of the linear combination are small) and (ii) the components that we want to move actually exist on respective sides (the coefficients have the right signs).

\begin{proof}
    Let $g = \gcd(A)$. If $p_{t-1}(c^t_x) = p_{t-1}(c^t_y)$, then $p = p_{t-1}$ is already $\C_t$-preserving and the lemma holds trivially. Hence, we may assume that $c^t_x$ and $c^t_y$ lie in different clusters. By symmetry, we may assume that (i)~$\size(c^t_x) \in \mult{g}$; denoting $x = \size(c^t_x)$, this means $g \mid x$ and $x \leq q$; and (ii)~$p_{t-1}(c^t_x) = 1$, i.e., component $c^t_x$ is in cluster $1$ at the beginning of step $t$.  

    We will create $p$ from $p_{t-1}$ by moving components between clusters so that $c^t_x$ changes its cluster, $c^t_y$ does not change its cluster, and in total, at most $O(q^2)$ elements change their clusters. This will ensure that $p$ is $\C_t$-preserving and $\dist(p_{t-1}, p) = O(q^2)$. 

    First, consider the case when $|A| = 1$, i.e., $A = \set{g}$. By the lemma assumption, $p_{t-1}$ is $(A, \sum_{i \in A} w(i))$-balanced. That is, cluster $0$ contains at least $\sum_{i \in A} w(i) = w(g) = 4 \cdot q^2 / g + 1 \geq q / g + 1 \geq x / g + 1$ components of size~$g$. At least $x / g$ of those components are different from~$c^t_y$. Hence, we can create $p$ from $p_{t-1}$ by swapping $c^t_x$ with these $x/g$ components of size~$g$ at the total cost of $x + (x/g) \cdot g = 2 \cdot x \leq 2 \cdot q$.

    Second, consider the case when $|A| \geq 2$. There is only one component, $c^t_z$, that is present in~$\C_t$, but not present in $\C_{t-1}$. Thus, by the lemma assumption, $\cnt_i(\C_{t-1}) \geq \cnt_i(\C_t) - 1 \geq 2 \cdot w(i)$ for each $i \in A$. Moreover, $\cnt_A(\C_{t-1}, p_{t-1}, y) \geq \sum_{i \in A} w(i)$ for each cluster $y \in \set{0, 1}$. Hence, by \cref{cla:balanced_partition} applied to $\C_{t-1}$, $p_{t-1}$, $A$, and $z \equiv w$, there exist two disjoint, non-empty sets $A_0, A_1$, such that $A = A_0 \uplus A_1$ and for each $y \in \set{0,1}$ and $i \in A_y$, it holds that $\cnt_i(\C_{t-1}, p_{t-1}, y) \geq w(i) = 4 \cdot q^2 / i + 1$. 
    
    By \cref{cla:number_theory} applied to sets $A_0$ and~$A_1$, there exist non-negative integers $r_i$, such that
    \begin{equation}
        \label{eq:balanced_move}
        \sum_{i \in A_0} r_i \cdot i = x + \sum_{i \in A_1} r_i \cdot i
    \end{equation}
    and $\sum_{i \in A_0} r_i \cdot i \leq 4 \cdot ( \max (A_0 \cup A_1 \cup \set{x}) )^2 \leq 4 \cdot q^2$, where the last inequality holds as $A \subseteq [q]$ and $x \leq q$. Thus, for any $i \in A_0$, we have $r_i \leq 4 \cdot q^2 / i$. As $x \geq 0$, we also have $\sum_{i \in A_1} r_i \cdot i \leq 4 \cdot q^2$, and thus the same bound on $r_i$ holds for each $i \in A_1$ as well.
    
    We create $p$ from $p_{t-1}$ by executing the following actions:
    \begin{itemize}
        \item For each $i \in A_0$, move $r_i$ components of size $i$ (other than $c^t_y$) from cluster $0$ to cluster~$1$. These components exist as $r_i + 1 \leq \cnt_i(\C_{t-1}, p_{t-1}, 0)$ for each $i \in A_0$.
        \item For each $i \in A_1$, move $r_i$ components of size $i$ (other than $c^t_x$) from cluster $1$ to cluster~$0$. These components exist as $r_i + 1 \leq \cnt_i(\C_{t-1}, p_{t-1}, 1)$ for each $i \in A_1$.
        \item Move component $c_x^t$ from cluster $1$ to cluster $0$.
    \end{itemize} 
    Observe that the resulting partition $p$ is $\C_t$-preserving as both $c^t_x$ and~$c^t_y$ end up in cluster~$0$, and \eqref{eq:balanced_move} ensures that the total number of elements in each cluster remains unchanged. Finally, the number of elements that change their cluster is $\sum_{i \in A_0} r_i \cdot i + \sum_{i \in A_1} r_i \cdot i + x \leq 2 \cdot (4 \cdot q^2) = O(q^2)$. 
\end{proof}


\section{Maintaining the core set of sizes}

In this section, we define an auxiliary sequence $g_1, g_2, \dots, g_t, \dots$, where $g_t \in [q] \cup \set{\infty}$, associated with the sequence of component sets $\C_1, \C_2, \dots$ within a single epoch. We call $g_t$ the \emph{GCD estimator} of step $t$ and $\mult{g_t}$ the \emph{core set of sizes} of step $t$.

The definition of $g_t$ balances two objectives: on the one hand, we want $g_t$ to be the greatest common divisor of those small component sizes for which sufficiently many components exist; on the other hand, we do not want $g_t$ to change too often. Therefore, $g_t$ is defined by the following iterative process:
\begin{equation}
    \label{eq:g_t}
    g_0 \define 1 
    \qquad \text{and} \qquad
    g_t \define \gcd \left( \mult{g_{t-1}} \cap \set{ i \suchthat \cnt_i(\C_t) \geq 2 \cdot w(i) + 1 } \right)
    \quad \text{for $t \geq 1$.}
\end{equation}
Note that as $\mult{g_{t-1}} \subseteq [q]$ for any $t$, this process ensures that $g_t \in [q] \cup \set{ \infty }$ for any $t$.

We say that a step $t$ is \emph{$g$-updating} if $g_t \neq g_{t-1}$ and otherwise we say it is \emph{$g$-preserving}. We now show that $g_t$ is non-decreasing, that in a~$g$-updating step the new value is a~multiple of the old one (and hence the core sets $\mult{g_t}$ only shrink over time), and that the number of $g$-updating steps is small. 

\begin{lemma}
\label{lem:g_structural}
    The number of $g$-updating steps within an epoch is at most $1 + \log q$.
    Furthermore, $\mult{g_t} \subseteq \mult{g_{t-1}}$ for each step $t$ of an epoch.
\end{lemma} 

\begin{proof}
    Fix an epoch $\Epoch$ and any step $t$ of $\Epoch$. Let $Z = \set{ i \suchthat \cnt_i(\C_t) \geq 2 \cdot w(i) + 1 }$. 
    By \cref{lem:gcd_property}, $\gcd(\mult{g_{t-1}}) \mid \gcd(\mult{g_{t-1}} \cap Z)$. 
    Note that $\gcd(\mult{g_{t-1}}) = g_{t-1}$ by \eqref{eq:mult_gcd}, and
    $\gcd(\mult{g_{t-1}} \cap Z) = g_t$ by \eqref{eq:g_t}. Thus, $g_{t-1} \mid g_t$ for any step $t$. 
    
    If $g_t < \infty$, then $g_{t-1} < \infty$, and consequently $\mult{g_t} \subseteq \mult{g_{t-1}}$ follows by the definition of~$\mult{\cdot}$. Otherwise $g_t = \infty$, in which case $\mult{g_t} = \emptyset \subseteq     \mult{g_{t-1}}$. Thus, the second part of the lemma follows.

    For the first part, let $\ell$ be the number of all $g$-updating steps within $\Epoch$; we denote them by $\tau(1), \tau(2), \dots, \tau(\ell)$. If $\ell = 0$, the bound holds trivially, and thus we assume $\ell \geq 1$. Let $\tau(0) = 0$. For any $i \in \set{0, \dots, \ell-1}$, it holds that $g_{\tau(i)} \mid g_{\tau(i+1)}$ and $g_{\tau(i)} \neq g_{\tau(i+1)}$, which implies $g_{\tau(i+1)} \geq 2 \cdot g_{\tau(i)}$. While it is possible that $g_{\tau(\ell)} = \infty$, we have $g_{\tau(\ell-1)} < \infty$: otherwise $g_{\tau(\ell-1)} \mid g_{\tau(\ell)}$ would enforce $g_{\tau(\ell)} = \infty = g_{\tau(\ell-1)}$. Thus, $g_{\tau(\ell-1)} \leq q$. As $g_{\tau(0)} = g_0 = 1$, we obtain $g_{\tau(\ell-1)} \geq 2^{\ell-1} \cdot g_{\tau(0)} = 2^{\ell-1}$, and therefore $2^{\ell-1} \leq q$, which concludes the first part of the lemma.
\end{proof}


\section{A subquadratic algorithm}

Our \textsc{Improved Component Based} algorithm (\ICB) is component-preserving. It splits an~epoch into two stages. The first stage is deterministic: with a slight ``rebalancing'' exception that we explain later, the components are remapped to minimize the cost of~changing the partition in a~single step. At a~carefully chosen step that we define later, \ICB switches to the second stage, which is randomized.


\subsection{Definition of ICB}

The pseudo-code of \ICB is given in \cref{alg:icb_1}; we describe it also below. It consists of three parts: the first one is executed in every step, and it is followed by the part corresponding to the current stage of an epoch.

\begin{algorithm}[t!]
\caption{A single epoch of \ICB\\
\textbf{Input:} initial partition $p_0$, sequence of component sets $\C_1, \C_2, \dots, \C_t, \dots$  \\
\textbf{Output:} sequence of partitions $p_1, p_2, \dots, p_t, \dots$, 
    where $p_t$ is $\C_t$-preserving unless the epoch terminates in step $t$ 
}
\label{alg:icb_1}
\begin{algorithmic}[1]
\algrenewcommand{\algorithmicrequire}{\textbf{In every step $t \geq 1$:}}
    \Require{}
    \If{$\P(\C_t) = \emptyset$} 
        \label{line:epoch_terminates_start}
        \Comment{no $\C_t$-preserving partition}
        \State{$p_t \gets p_{t-1}$}
        \State{\textbf{terminate} the current epoch} 
        \label{line:epoch_terminates_end}
    \EndIf
\algrenewcommand{\algorithmicrequire}{\textbf{Additionally, in every step of the first stage:}}
    \Require{}
    \State{Compute $g_t$ according to \eqref{eq:g_t}}
        \label{line:compute_g_t}
    \State{$p^*_t \gets \arg\min_p \{ \dist(p_{t-1}, p) : p \in \P(\C_t) \}$} 
        \Comment{pick closest $\C_t$-preserving candidate}
        \label{line:candidate_solution}    
    \If{$p^*_t \in \P(\C_t, \mult{g_t}, d)$}
    \label{line:choosing_p_t_start}
    \label{line:rebalance_condition}
        \State{$p_t \gets p^*_t$}
    \ElsIf{$\P(\C_t, \mult{g_t}, 2d) \neq \emptyset$}
        \State{$p_t \gets \text{any partition from } \P(\C_t, \mult{g_t}, 2d)$} 
            \Comment{rebalancing}
            \label{line:choosing_p_t_end}    
    \Else{}
        \label{line:terminate_first_stage_start}
        \Comment{no $\C_t$-preserving $(\mult{g_t},2d)$-balanced partition}
        \State{$p_t \gets p^*_t$}
        \State{\textbf{terminate} the first stage of the current epoch}    
        \label{line:terminate_first_stage_end}
    \EndIf
\algrenewcommand{\algorithmicrequire}{\textbf{Additionally, in every step of the second stage:}}
    \Require{}
    \If{$p_{t-1} \in \P(\C_t)$}
        \label{line:second_stage_start}
        \State{$p_t \gets p_{t-1}$}
    \Else{}
        \State{$p_t \gets \text{a partition chosen uniformly at random from } \P(\C_t)$}
            \Comment{re-sampling}
        \label{line:second_stage_end}
    \EndIf
\end{algorithmic}
\end{algorithm}


\paragraph{Terminating an epoch.}

In \crefrange{line:epoch_terminates_start}{line:epoch_terminates_end}, executed in both stages, \ICB verifies whether a~$\C_t$-preserving partition exists, and terminates the epoch without changing the current partition otherwise. In particular, in all the remaining lines, $\P(\C_t) \neq \emptyset$.


\paragraph{Computing the GCD estimator.}

In \cref{line:compute_g_t}, \ICB updates the GCD estimator $g_t$ according to~\eqref{eq:g_t}. Note that $g_t$ is used only within the first stage of an epoch.


\paragraph{Choosing a new partition.}

In \cref{line:candidate_solution}, \ICB computes a \emph{candidate partition} $p^*_t$ as a~$\C_t$-preserving partition closest to~$p_{t-1}$. If $p^*_t$ is $(\mult{g_t}, d)$-balanced, \ICB simply outputs $p_t = p^*_t$. Otherwise, it discards $p^*_t$ and picks any $(\mult{g_t}, 2d)$-balanced partition as $p_t$ (\cref{line:choosing_p_t_end}). We call such an action \emph{rebalancing}; we later show that it occurs rarely, i.e., in most cases $p_t = p^*_t$.


\paragraph{Triggering the second stage.}

Rebalancing is possible only when a $\C_t$-preserving $(\mult{g_t},2d)$-balanced partition exists. If it does not, \ICB outputs $p_t = p^*_t$, terminates the first stage (\crefrange{line:terminate_first_stage_start}{line:terminate_first_stage_end}), and switches to the second stage of an epoch in the next step.


\paragraph{The second stage.}

The second stage is much simpler (\crefrange{line:second_stage_start}{line:second_stage_end}): \ICB keeps its partition as long as it remains $\C_t$-preserving, and once it stops being one, \ICB replaces it by a partition chosen uniformly at random from~$\P(\C_t)$.


\section{Analysis roadmap}

In this section, we describe the framework of our analysis in a top-down approach, listing the necessary lemmas that will be proven in the next sections, and showing that their combination yields the desired competitiveness bound.

\cref{lem:epoch_cost_only} allows us to focus only on the cost of \ICB in a single epoch $\Epoch$. We denote its two stages by $\Epoch_1$ and $\Epoch_2$; the second stage is empty if the epoch terminates in \crefrange{line:epoch_terminates_start}{line:epoch_terminates_end} while \ICB is still in the first stage. We identify $\Epoch_1$ and $\Epoch_2$ with the sets of the corresponding steps. In particular, we use $T$ as the number of steps in the first stage, i.e., $\Epoch_1 = [T]$. If $\Epoch_2$ is non-empty, then $T$ is the step in which \ICB executes \crefrange{line:terminate_first_stage_start}{line:terminate_first_stage_end}, and thus $p_T = p^*_T$.

In the following two sections, we prove the bounds below on the cost of \ICB in the two stages of an epoch. The proofs are deferred to \cref{sec:first_stage,sec:second_stage}.

\begin{restatable}{lemma}{stageOneCost}
\label{lem:stage_one_cost}
    For any epoch $\Epoch$, the cost of \ICB in the first stage of an epoch is $\ICB(\Epoch_1) = O(n^2 \cdot q^2 / d + n \cdot q^2 \cdot \log q + n^2 / q)$.
\end{restatable}

\begin{restatable}{lemma}{stageTwoCost}
\label{lem:stage_two_cost}
    For any epoch $\Epoch$, the expected cost of \ICB in the second stage of an epoch (assuming the second stage is present) is $\E[\ICB(\Epoch_2)] = O(n \cdot (d \cdot \log (n/d) + q^2 \cdot \log q + n / q))$.
\end{restatable}

\begin{theorem}
\label{thm:competitive_ratio}
    \ICB is $O(n^{7/4} \cdot \log^{1/4} n)$-competitive for the online bisection problem.
\end{theorem}

\begin{proof}
    We set $q = \lceil n^{1/4} / \log^{1/4} n \rceil$ and $d = \lceil n^{3/4} / \log^{3/4} n \rceil$. 

    Recall that we defined $w(i) = 4 \cdot q^2 / i + 1$ for each $i \in [q]$, and thus $\sum_{i \in [q]} w(i) = O(q^2 \cdot \log q)$. Hence, the setting of $q$ and $d$ ensures that $d \geq \sum_{i \in [q]} (2 \cdot w(i) + 3)$ and $2 \cdot d \leq n$ (as required by \eqref{eq:parameter_values_relations}) for sufficiently large $n$.
    
    Applying \cref{lem:stage_one_cost} and \cref{lem:stage_two_cost}, and using $\log q \leq \log n$ and $\log (n/d) \leq \log n$, we obtain
    \begin{align*}
        \E[\ICB(\Epoch)]
        & = O \left(
        n^2 \cdot q^2 / d + n \cdot q^2 \cdot \log n + n^2 / q +
        n \cdot d \cdot \log n
        \right) \\
        & = O \left( n^{7/4} \cdot \log^{1/4} n + n^{3/2} \cdot \log^{1/2} n + n^{7/4} \cdot \log^{1/4} n + n^{7/4} \cdot \log^{1/4} n \right) \\
        & = O \left(n^{7/4} \cdot \log^{1/4} n \right).
    \end{align*}
    The theorem follows immediately by \cref{lem:epoch_cost_only}.
\end{proof}


\section{Analysis: the first stage of ICB}
\label{sec:first_stage}

Recall that in a~single step $t$, \ICB pays at most~$1$ for serving the request and $\dist(p_{t-1}, p_t)$ for changing the partition. We may upper-bound the latter term by $\dist(p_{t-1}, p^*_t) + \dist(p^*_t, p_t)$; we call the first summand \emph{switching cost} and the second one \emph{rebalancing cost}. We later show that rebalancing occurs rarely, and thus its overall cost is small. We define  
\begin{align*}
    \ICBss(t) &\define 1+ \dist(p_{t-1}, p^*_t), \\
    \ICBrb(t) &\define \dist(p^*_t, p_t).
\end{align*}
If the epoch terminates in step $t$ of the first stage (in \crefrange{line:epoch_terminates_start}{line:epoch_terminates_end}), then $p^*_t$ is not computed by \ICB, and for the purpose of the analysis, we set $p^*_t = p_{t-1} = p_t$. By the triangle inequality, $\ICB(t) \leq \ICBss(t) + \ICBrb(t)$. For a set $S$ of steps, we write $\ICB(S) \define \sum_{t \in S} \ICB(t)$, and analogously for $\ICBss$ and $\ICBrb$.


\subsection{Bounding serving and switching costs}

We first estimate $\ICBss(\Epoch_1)$. 

Recall that $c^t_x$ and $c^t_y$ are the components merged in step $t$. To bound the switching cost, we distinguish between regular and irregular steps. In regular ones, at least one merged component is small and its size is divisible by $g_{t-1}$. 

\begin{definition}
    A step~$t$ is \emph{regular} if $\size(c^t_x) \in \mult{g_{t-1}}$ 
    or $\size(c^t_y) \in \mult{g_{t-1}}$, and \emph{irregular} otherwise.
\end{definition}

We now argue that the switching and serving cost in regular steps is $o(n)$. To this end, we observe that \cref{alg:icb_1} executed in step $t-1$ ensures (by running rebalancing if necessary) that $p_{t-1}$ is a $(\mult{g_{t-1}},d)$-balanced partition from $\P(\C_{t-1})$, i.e., both clusters of partition $p_{t-1}$ contain at least $d$ small components of sizes divisible by $g_{t-1}$. For $t = 1$, the same property holds trivially: $\C_0$~consists of $n$ singleton components, $g_0 = 1$, and hence $\cnt_{\mult{g_0}}(\C_0,p_0,y) = n/2 \geq d$ for each cluster~$y$ (as $2 \cdot d \leq n$). This property becomes useful at the beginning of step~$t$: using number-theoretic arguments, we may bound the number of components that need to be moved between clusters, so that eventually $c^t_x$ and $c^t_y$ end up in the same cluster. 

\begin{claim}
\label{cla:infinite_g}
    For any step $t$ of $\Epoch_1$, it holds that $g_{t-1} < \infty$. 
\end{claim}

\begin{proof}
    As $g_0 = 1$, the claim holds trivially for $t = 1$. Hence, we assume that $t > 1$. Suppose that $g_{t-1} = \infty$. Then, $\mult{g_{t-1}} = \emptyset$, and therefore both $\P(\C_{t-1},\mult{g_{t-1}},d)$ and $\P(\C_{t-1},\mult{g_{t-1}},2d)$ are empty. Thus, in step $t-1$, \ICB would reach \crefrange{line:terminate_first_stage_start}{line:terminate_first_stage_end} and terminate $\Epoch_1$ already in step $t-1$.
\end{proof}

\begin{lemma}
\label{lem:regular_cost}
    For each regular $g$-preserving step $t$ of $\Epoch_1$, $\ICBss(t) = O(q^2)$.
\end{lemma}

\begin{proof}
    Recall that $\ICBss(t) = 1 + \dist(p_{t-1}, p^*_t)$, and $p^*_t$ is the partition from $\P(\C_t)$ closest to $p_{t-1}$. Thus it suffices to show that there exists a partition $p \in \P(\C_t)$ such that $\dist(p_{t-1}, p) = O(q^2)$. We prove it by showing that the set $A$ defined as 
    \[ 
        A \define \mult{g_{t-1}} \cap Z \qquad \text{where} \qquad Z \define \set{ i \suchthat \cnt_i(\C_t) \geq 2 \cdot w(i) + 1}
    \]
    satisfies the properties of \cref{lem:regular_merge}. Note that $A \subseteq \mult{g_{t-1}} \subseteq [q]$. Moreover, $A \neq \emptyset$: otherwise $g_t = \gcd(A) = \infty$, and as step $t$ is $g$-preserving, also $g_{t-1} = \infty$, contradicting \cref{cla:infinite_g}.

    The first property of \cref{lem:regular_merge} follows immediately as $A \subseteq Z$.

    To show the second property of \cref{lem:regular_merge}, we first observe that by~\eqref{eq:g_t}, $g_t = \gcd(\mult{g_{t-1}} \cap Z)$, and thus $\gcd(A) = g_t$. As the step is $g$-preserving, $g_t = g_{t-1}$, and thus $\gcd(A) = g_{t-1}$. As the step is regular, $\size(c^t_x) \in \mult{g_{t-1}} = \mult{\gcd(A)}$ or $\size(c^t_y) \in \mult{g_{t-1}} = \mult{\gcd(A)}$.

    Finally, for the third property of \cref{lem:regular_merge}, we fix a cluster $y \in \{0, 1\}$ and we lower-bound $\cnt_A(\C_{t-1}, p_{t-1}, y)$. To this end, we first observe that \cref{alg:icb_1} executed in step~$t-1$ ensures that $\cnt_{\mult{g_{t-1}}} (\C_{t-1}, p_{t-1}, y) \geq d$. 
    Moreover, for a component size $i \in \mult{g_{t-1}} \setminus Z$, we have $\cnt_i(\C_t) < 2 \cdot w(i) + 1$. As there are only two components, $c^t_x$ and $c_y^t$, that are present in~$\C_{t-1}$ but not present in~$\C_t$, we have $\cnt_i(\C_{t-1}) < 2 \cdot w(i) + 3$ for such $i$. Hence, 
    \begin{align*}
        \cnt_A(\C_{t-1}, p_{t-1}, y) 
        & = \cnt_{\mult{g_{t-1}}}(\C_{t-1}, p_{t-1}, y) 
        - \cnt_{\mult{g_{t-1}} \setminus Z}(\C_{t-1}, p_{t-1}, y)  \\
        & \geq d - \sum_{i \in \mult{g_{t-1}} \setminus Z} (2 \cdot w(i) + 3) \\
        & \geq \sum_{i \in [q]} (2 \cdot w(i) + 3) - 
        \sum_{i \in \mult{g_{t-1}} \setminus Z} (2 \cdot w(i) + 3) \\
        & \geq \sum_{i \in A} (2 \cdot w(i) + 3),
    \end{align*}
    where the last inequality follows as $A$ and $\mult{g_{t-1}} \setminus Z$ are disjoint subsets of~$[q]$. As the bound above holds for each cluster $y \in \set{0, 1}$, partition $p_{t-1}$ is $(A, \sum_{i \in A} (2 \cdot w(i) + 3))$-balanced, and hence also $(A, \sum_{i \in A} w(i))$-balanced, satisfying the third property of \cref{lem:regular_merge}. This concludes the proof.
\end{proof}


\subsection{Bounding the number of irregular steps}
\label{sec:irregular_steps}

Next, we argue that there are $o(n)$ irregular steps, and we also bound the number of components whose sizes are either large or not divisible by $g_t$.

To bound the number of irregular steps, we trace the evolution of components. Recall that $\C_t = \C_{t-1} \cup \, \{c^t_z\} \setminus \{c^t_x,c^t_y\}$, i.e., components $c^t_x$ and $c^t_y$ are merged in step~$t$ into component~$c^t_z$. We say that components $c^t_x$ and $c^t_y$ are \emph{destroyed} in step $t$ and component $c^t_z$ is \emph{created} in step $t$. We extend these notions also to the (singleton) components of $\C_0$, where we say that they are created in step $0$, and to components of $\C_T$, where we say that they are destroyed in step~$T+1$. 

We now fix a small component $c$ created at time $a$ and destroyed at time $b$. Note that $0 \leq a < b \leq T+1$. 
By \cref{lem:g_structural}, $\mult{g_{b-1}} \subseteq \mult{g_a}$. We say that the component $c$ is 
\begin{itemize}
    \item \emph{typical} if $\size(c) \in \mult{g_{b-1}}$,
    \item \emph{mixed} if $\size(c) \in \mult{g_a} \setminus \mult{g_{b-1}}$,
    \item \emph{atypical} if $\size(c) \notin \mult{g_a}$.
\end{itemize}
That is, each component is either large, typical, atypical, or mixed. In particular, in a regular merge, at least one of the merged components is typical.

\begin{claim}
\label{cla:two_typical_components}
    Assume step $t$ is $g$-preserving and both $c_x^t$ and $c_y^t$ are typical. Then, $c_z^t$ is not atypical.
\end{claim}

\begin{proof}
    As components are typical, $\size(c_x^t) \in \mult{g_{t-1}}$ and $\size(c_y^t) \in \mult{g_{t-1}}$, and therefore $g_{t-1} \mid \size(c_x^t)$ and $g_{t-1} \mid \size(c_y^t)$. As $\size(c_z^t) = \size(c_x^t) + \size(c_y^t)$, we have $g_{t-1} \mid \size(c_z^t)$. Finally, as step $t$ is $g$-preserving, $g_t = g_{t-1}$, and hence $g_t \mid \size(c_z^t)$. If $c_z^t$ is large, then the claim follows immediately. If $c_z^t$ is small, then we have $\size(c_z^t) \in \mult{g_t}$, and thus $c_z^t$~cannot be atypical.
\end{proof}


\paragraph{Merge forest.}

It is convenient to consider the following \emph{merge forest}~$\F$, whose nodes correspond to all components that appear within $\Epoch_1$. We connect these nodes by edges in a~natural manner: the leaves of~$\F$ correspond to initial singleton components of~$\C_0$, and each internal node of $\F$ corresponds to a~component created by merging its children components. We say that an internal node of $\F$ is large/typical/atypical/mixed if the corresponding component is of such type.


\paragraph{Types of irregular merges.}

To upper-bound the number of irregular merges, we subdivide them into three types.
\begin{itemize}
    \item \emph{All-large irregular merges}: both merged components are large (and the resulting component is clearly large as well). 
    \item \emph{Mixed-resulting irregular merges}: the created component is mixed.
    \item \emph{Ordinary irregular merges}: all other irregular merges.
\end{itemize}
More generally, and also for regular steps, we call a step \emph{mixed-resulting} if the component created in this step is mixed. We bound the number of these merges separately in the following claims.

\begin{claim}
\label{cla:F_large}
    $\F$ contains at most $n / q$ internal nodes both of whose children are large.
\end{claim}

\begin{proof}
    Let $L$ be the set of large nodes of $\F$. Clearly, $L$ is ``upward-closed'', i.e., if $L$ contains a~node, then it also contains its parent. Let $\F_L$ be the sub-forest of $\F$ induced by nodes from~$L$. We partition $L$ into three sets: $L_0$, $L_1$ and $L_2$, where a component from $L_i$ has exactly $i$ children in $\F_L$. We need to show that $|L_2| \leq n / q$.

    As $L_2$ contains the branching internal nodes of $\F_L$ and $L_0$ contains the leaves of $\F_L$, we have $|L_2| < |L_0|$. All components from $L_0$ are large, i.e., each of them consists of at least $q+1$ elements. Fix any two components from $L_0$. As they are leaves of $\F_L$, they are not in the ancestor-descendant relation in~$\F_L$, and, by the upward-closedness of~$L$, also not in~$\F$; hence the sets of their elements are disjoint. Thus, all components of $L_0$ are disjoint, which implies $|L_0| \cdot (q+1) \leq n$. Summing up, $|L_2| < |L_0| \leq n / (q+1)$.
\end{proof}

\begin{claim}
\label{cla:F_mixed}
    $\F$ contains $O(q^2 \cdot \log q)$ mixed nodes.
\end{claim}

\begin{proof}
    Fix any mixed node corresponding to component $c$ that is created at step $a$ and destroyed at step $b > a$. By \cref{lem:g_structural}, $\mult{g_a} \supseteq \mult{g_{a+1}} \supseteq \ldots \supseteq \mult{g_{b-1}}$. As $\size(c) \in \mult{g_a}$ and $\size(c) \notin \mult{g_{b-1}}$, there exists a step $t \in [a+1,b-1]$, such that $\size(c) \in \mult{g_{t-1}} \setminus \mult{g_t}$. We say that component $c$ is \emph{$t$-mixed}.

    We now upper-bound the number of $t$-mixed components. Fix $t \in [T]$ and a component size $j \in \mult{g_{t-1}} \setminus \mult{g_t}$. Let $Z = \mult{g_{t-1}} \cap \set{ i \suchthat \cnt_i(\C_t) \geq 2 \cdot w(i) + 1 }$. Suppose for a~contradiction that $\cnt_j(\C_t) \geq 2 \cdot w(j) + 1$. Then, $j \in Z$. On the other hand, $g_t = \gcd(Z)$, and thus $g_t \mid j$. However, as $j \in [q]$, we would then have $j \in \mult{g_t}$, a contradiction.
    Thus, we have that $\cnt_j(\C_t) < 2 \cdot w(j) + 1$. Any $t$-mixed component belongs to $\C_t$, and hence the number of $t$-mixed components is at most $\sum_{i \in \mult{g_{t-1}} \setminus \mult{g_t}} (2 \cdot w(i) + 1)$.
    
    Any mixed node is $t$-mixed for some step $t \in [T]$ and, as the sets $\mult{g_t}$ are nested by \cref{lem:g_structural}, the sums below telescope. Hence, the total number of all mixed nodes is at most $\sum_{t \in [T]} \sum_{i \in \mult{g_{t-1}} \setminus \mult{g_t}} (2 \cdot w(i) + 1) = \sum_{i \in \mult{g_0} \setminus \mult{g_T}} (2 \cdot w(i) + 1) \leq \sum_{i \in [q]} (2 \cdot w(i) + 1)$. Plugging $w(i) = 4 \cdot q^2 / i + 1$ yields the~claim.
\end{proof}

It remains to bound the number of ordinary irregular merges. To this end, we define the following amounts (for any $t \geq 0$).
\begin{itemize}
    \item $a_t$ is the number of components in $\C_t$ that are atypical or mixed.
    \item $I_t$ is the number of ordinary irregular merges in steps $1, 2, \dots, t$.
\end{itemize}

\begin{claim}
\label{cla:irregular_induction}
    For any step $t \geq 0$, it holds that $a_t = O(q^2 \cdot \log q)$ and $I_t = O(q^2 \cdot \log q)$. 
\end{claim}

\begin{proof}
    Let $\Delta_t a = a_t - a_{t-1}$ and $\Delta_t I = I_t - I_{t-1}$. Clearly, $\Delta_t a \in \{-2, -1, 0, 1\}$ and $\Delta_t I \in \{0, 1\}$ for any step $t \geq 1$. We analyze the value of $\Delta_t a + \Delta_t I$ depending on the type of step~$t$, considering the following cases.

    \begin{itemize}
        \item Step $t$ is $g$-updating. Then, $\Delta_t a \leq 1$ and $\Delta_t I \leq 1$.
        \item Step $t$ is $g$-preserving and all-large irregular. Then, $\Delta_t I = 0$ and $\Delta_t a = 0$.
        \item Step $t$ is $g$-preserving and mixed-resulting (irregular or regular). Then, $\Delta_t I = 0$, and $\Delta_t a \leq 1$.
        \item Step $t$ is $g$-preserving and ordinary irregular, i.e., $\Delta_t I = 1$. We show that $\Delta_t a \leq -1$. In this case at most one of the merged components is large and none is typical.
            \begin{itemize} 
            \item If both merged components, $c_x^t$ and $c_y^t$, are atypical or mixed, then we immediately obtain $\Delta_t a \leq -1$. 
            \item Only one of merged components, say $c_x^t$, is atypical or mixed. Then the second one,~$c_y^t$, must be large, and thus $c_z^t$ is large as well. Hence, $\Delta_t a \leq -1$.
            \end{itemize}
        \item Step $t$ is $g$-preserving and regular, but not mixed-resulting. Then, at least one of the merged components, say $c_x^t$, is typical, $\Delta_t I = 0$, and we show that $\Delta_t a \leq 0$. 
            \begin{itemize}
            \item If $c_y^t$ is also typical, then $c_z^t$ cannot be   atypical by \cref{cla:two_typical_components}. As we assumed $c_z^t$ is not mixed, it must be either large or typical. Hence, $\Delta_t a = 0$.
            \item If $c_y^t$ is atypical or mixed, then $\Delta_t a \leq 0$.
            \item If $c_y^t$ is large, then $c_z^t$ is large as well, and then $\Delta_t a = 0$.
            \end{itemize}
        \end{itemize}

    We note that $a_0 = I_0 = 0$. By the case analysis above, $\Delta_t a + \Delta_t I \leq 2$ when step $t$ is $g$-updating, $\Delta_t a + \Delta_t I \leq 1$ when step $t$ is $g$-preserving and mixed-resulting, and $\Delta_t a + \Delta_t I \leq 0$ in all remaining cases. Hence, by summing over all steps up to~$t$, we obtain that $a_t + I_t$ is upper-bounded by twice the number of steps that are either $g$-updating or mixed-resulting. Their number is at most $O(q^2 \cdot \log q)$ by \cref{lem:g_structural} and \cref{cla:F_mixed}. The claim follows as both $a_t$ and $I_t$ are non-negative.
\end{proof}

\begin{lemma}
\label{lem:irregular_steps_count}
    There are at most $O(q^2 \cdot \log q + n / q)$ irregular steps in $\Epoch_1$. Moreover, at any time~$t$~of~$\Epoch_1$, $\cnt_{[n] \setminus \mult{g_t}}(\C_t) = O(q^2 \cdot \log q + n / q)$.
\end{lemma}
    
\begin{proof}
    There are at most $n / q$ all-large irregular steps by \cref{cla:F_large}, at most $O(q^2 \cdot \log q)$ mixed-resulting irregular steps by \cref{cla:F_mixed}, and $I_T = O(q^2 \cdot \log q)$ ordinary irregular steps by \cref{cla:irregular_induction}. This shows the first part of the lemma.

    For the second part, fix any step $t$. Observe that components from $\C_t$ whose sizes are from $[n] \setminus \mult{g_t}$ are not typical, i.e., they must be either large, atypical, or mixed. Trivially, there are at most $n / (q+1)$ large components, and the number of atypical and mixed components is $a_t = O(q^2 \cdot \log q)$ by \cref{cla:irregular_induction}.
\end{proof}


\subsection{Bounding rebalancing costs}

In this section, we bound the total rebalancing cost by upper-bounding the number of rebalancing events. To this end, we focus on chunks of steps between two consecutive rebalancing events. We show that each such chunk either contains a rare step (an irregular or a $g$-updating one) or consists of many steps. As rare steps are few and the total number of steps is at most $n$, the number of chunks is small.

\begin{claim}
    \label{cla:between_rebalancing}
    Let $a$ and $b$ be two consecutive steps where rebalancing is executed. 
    If $g_a = g_b$ and $b - a < d/3$, then
    $\dist(p_a, p^*_b) \geq d/3$.
\end{claim}

\begin{proof}
    Let $g = g_a = g_b$. As~rebalancing was triggered in step $b$, we have $p^*_b \notin \P(\C_b,\mult{g},d)$. That is, partition $p^*_b$ has fewer than $d$~components from $\C_b$ of sizes from $\mult{g}$ in one of the clusters (say, in cluster~$0$). Observe that $\C_a$ can be obtained from $\C_b$ by going back in time and reversing component merges, i.e., performing $b-a$ splits of components. Each such split may create two extra components of size from $\mult{g}$. Hence, partition $p^*_b$ keeps fewer than $d + 2 \cdot (b-a) < d + (2/3) \cdot d$ components of~$\C_a$ of sizes from~$\mult{g}$ in cluster $0$. 
    
    Due to rebalancing in step $a$, we have $p_a \in \P(\C_a,\mult{g},2d)$, i.e., $p_a$ keeps at least $2 d$ components of~$\C_a$ of sizes from~$\mult{g}$ in cluster $0$. Hence, more than $2 d - (d + (2/3) \cdot d) = d/3$ components of~$\C_a$ of sizes from~$\mult{g}$ are placed in cluster $0$ by $p_a$ and in cluster $1$ by $p^*_b$. As each of them contains at least one element, we obtain $\dist(p_a, p^*_b) \geq d/3$.
\end{proof}

\begin{claim}
    \label{cla:between_rebalancing_2}
    Let $a$ and $b$ be two consecutive steps where rebalancing is executed. Assume that all steps of $\set{a+1, \dots, b}$ are regular and $g$-preserving. Then, $b-a = \Omega(d/q^2)$.
\end{claim}

\begin{proof}
    Assume that $b-a < d/3$ as otherwise the claim follows trivially. As all steps of $\set{a+1, \dots, b}$ are $g$-preserving, $g_a = g_b$. Note that for any step $t \in \{ a+1, \dots, b-1\}$, there is no rebalancing in $t$, and thus $p_t = p_t^*$. Hence,
    \begin{align*}
        d/3
        & \leq \textstyle \dist(p_a, p^*_b)
            && \text{(by \cref{cla:between_rebalancing})} \\
        & \leq \textstyle \sum_{t=a+1}^{b-1} \dist(p_{t-1}, p_t) + \dist(p_{b-1}, p^*_b)
            && \text{(by the triangle inequality)} \\
        & = \textstyle \sum_{t=a+1}^b \dist(p_{t-1}, p^*_t)
            && \text{(as $p_t = p^*_t$ for all $a + 1 \leq t \leq b - 1$)} \\
        & = \textstyle \sum_{t=a+1}^{b} (\ICBss(t) - 1)
            && \text{(by the definition of $\ICBss$)} \\
        & \leq \textstyle (b-a) \cdot O(q^2),
            && \text{(by \cref{lem:regular_cost})}
    \end{align*}
    which concludes the proof.
\end{proof}

\begin{lemma}
\label{lem:rebalancing_overhead}
    It holds that $\ICBrb(\Epoch_1) = O(n^2 \cdot q^2 / d + n \cdot q^2 \cdot \log q + n^2 / q)$.
\end{lemma}

\begin{proof}
    Let $\ell$ be the number of rebalancing events within $\Epoch_1$. As the cost of a single rebalancing event is at most $n$, we have $\ICBrb(\Epoch_1) \leq \ell \cdot n$, and thus it suffices to show that $\ell = O(n \cdot q^2 / d + q^2 \cdot \log q + n / q)$.

    The rebalancing events partition $\Epoch_1$ into $\ell+1$ disjoint chunks; we neglect the first chunk and the last one. Each of the $\ell-1$ remaining chunks is of the form $\set{a+1, \dots, b}$, where $a$ and $b$ are two consecutive steps in which rebalancing is executed. We consider two types of chunks.
    \begin{itemize}
        \item \emph{Chunks containing a $g$-updating step or an irregular step.}
            There are at most $1 + \log q +  O(q^2 \cdot \log q + n / q)$ such chunks, by \cref{lem:g_structural} and \cref{lem:irregular_steps_count}.
        \item \emph{Chunks containing only regular $g$-preserving steps.}
            Fix such a chunk $\set{a+1, \dots, b}$. By \cref{cla:between_rebalancing_2}, 
            $b - a = \Omega(d/q^2)$. As the total number of steps in all such chunks is smaller than $n$ (cf.~\cref{lem:partition_preserving_trivial}), there are at most $O(n / (d/q^2)) = O(n \cdot q^2 / d)$ such chunks.
    \end{itemize}
    Summing up, $\ell = O(q^2 \cdot \log q + n / q + n \cdot q^2 / d)$, which concludes the proof. 
\end{proof}


\subsection{Estimating the total cost}

We may now combine the bounds presented above to prove \cref{lem:stage_one_cost}, which we restate below. 

\stageOneCost*

\begin{proof}
    Let $T$ be the number of steps in $\Epoch_1$. Let $R \subseteq [T]$ be the set of regular steps of~$\Epoch_1$ that are not $g$-updating. Then each step from $[T] \setminus R$ is either irregular or \mbox{$g$-updating}. By \cref{lem:g_structural} and \cref{lem:irregular_steps_count},
    \begin{equation}
    \label{eq:not-regular-bound}
        |[T] \setminus R| \leq (1+\log q) + O(q^2 \cdot \log q + n / q) = O(q^2 \cdot \log q + n/q).
    \end{equation}
    This allows us to upper-bound the serving and switching cost of \ICB in $\Epoch_1$ as
    \begin{align*}
        \ICBss(\Epoch_1) 
        & = \sum_{t \in R} \ICBss(t) + \sum_{t \in [T] \setminus R} \ICBss(t) \\*
        & \leq \sum_{t \in R} O(q^2) + \sum_{t \in [T] \setminus R} (n+1) 
            && \text{(by~\cref{lem:regular_cost} and \cref{lem:partition_preserving_trivial})} \\*
        & = |R| \cdot O(q^2) + O(q^2 \cdot \log q + n/q) \cdot (n+1) 
            && \text{(by \eqref{eq:not-regular-bound})} \\*
        & = O(n \cdot (q^2 \cdot \log q + n/q)).
            && \text{(as $|R| \leq T \leq n-1$)}
    \end{align*}
    The total cost in $\Epoch_1$ (including rebalancing) is then
    \begin{align*}
        \ICB(\Epoch_1) 
            & \leq \ICBss(\Epoch_1) + \ICBrb(\Epoch_1) \\
            & = O(n \cdot q^2 \cdot \log q + n^2/q) + O(n^2 \cdot q^2 / d + n \cdot q^2 \cdot \log q + n^2 / q)
                && \text{(by~\cref{lem:rebalancing_overhead})} \\
            & = O \left( n^2 \cdot q^2 / d + n \cdot q^2 \cdot \log q + n^2 / q \right).
                && \qedhere
    \end{align*}
\end{proof}


\section{Analysis: the second stage of ICB}
\label{sec:second_stage}

Now we proceed to analyze the second stage of \ICB. We assume that $\Epoch_2$ is non-empty as otherwise the associated cost is trivially zero. That is, \ICB executed \crefrange{line:terminate_first_stage_start}{line:terminate_first_stage_end} in step $T$, which happens only when $\P(\C_T, \mult{g_T}, 2d) = \emptyset$.

We start with a combinatorial counting lemma, which estimates the cardinality of~$\P(\C_T)$. Next, we show that in the second stage, \ICB changes its partition $O(\log |\P(\C_T)|)$ times in expectation, each time paying at most $n+1$.

\begin{lemma}
\label{lem:P_C_T_bound}
    If the second stage is present, $|\P(\C_T)| = \exp(O(d \cdot \log (n/d) + q^2 \cdot \log q + n / q))$.
\end{lemma}

\begin{proof}
    We partition components of $\C_T$ into two sets: $A$ containing components of sizes from~$\mult{g_T}$ and $B$ containing remaining components. Recall that $|\P(\C_T)|$ is the number of ways we can feasibly assign components of $\C_T$ to two clusters. We separately bound the number of ways of assigning components from $A$ and $B$ to two clusters.
    
    As \ICB terminated the first stage in step $T$, we have $\P(\C_T, \mult{g_T}, 2d) = \emptyset$, i.e., each feasible mapping places fewer than $2 \cdot d$ components from $A$ on at least one side. Thus, the overall number of ways of partitioning the components of $A$ among two clusters is at most 
    \[ 
        \sum_{i=0}^{2 d - 1} 2 \cdot \binom{|A|}{i} 
            \leq 2 \cdot \sum_{i=0}^{2 d} \binom{n}{i}  
            \leq 2 \cdot \left( \frac{e \cdot n}{2 \cdot d} \right)^{2 d} 
            = \exp(O(d \cdot \log (n/d))).
    \]

    On the other hand, the number of ways of assigning components from $B$ to two clusters is at most $2^{|B|}$. By the definition, $|B| = \cnt_{[n] \setminus \mult{g_T}}(\C_T)$ which is at most $O(q^2 \cdot \log q + n / q)$ by \cref{lem:irregular_steps_count}. 

    Hence
    \begin{align*}
        |\P(\C_T)|
        & \leq \exp(O(d \cdot \log (n/d))) \cdot 2^{|B|} \\*
        & = \exp(O(d \cdot \log (n/d) + |B|)) \\*
        & = \exp(O(d \cdot \log (n/d) + q^2 \cdot \log q + n / q)). 
            && \qedhere
    \end{align*}
\end{proof}

It remains to bound the expected number of times \ICB changes its partition in $\Epoch_2$. Its random decisions are captured by the claim below. The properties of the induced random process are well-known; for instance, the same process underlies the classic randomized algorithm for the metrical task system problem on a~uniform metric~\cite{BoLiSa92}. We include a proof for completeness in \cref{sec:mts}.

\begin{claim}
    \label{cla:random_survivor}
    Consider a process on a~finite non-empty set $S_0$ of states, which in round $j \geq 1$ replaces the current set $S_{j-1}$ of \emph{alive} states by an arbitrary non-empty subset $S_j \subseteq S_{j-1}$. An algorithm starts at an~arbitrary state of $S_0$, and, at the end of round $j$, if its current state is not in $S_j$, it \emph{moves} to a state chosen uniformly at random from $S_j$. The expected number of moves is at most $1 + \ln |S_0|$. 
\end{claim}

Combining these two ingredients, we obtain the bound on the cost of the second stage, which we restate below.

\stageTwoCost*

\begin{proof}
    Recall that $p_T \in \P(\C_T)$. By \crefrange{line:second_stage_start}{line:second_stage_end}, in a step~$t$ of~$\Epoch_2$, \ICB changes its partition only when its current partition is not $\C_t$-preserving, and in such a case it chooses a new partition $p_t$ uniformly at random from $\P(\C_t)$. As $\C_t$ is coarser than $\C_{t-1}$, we have $\P(\C_t) \subseteq \P(\C_{t-1})$. Thus, \cref{cla:random_survivor} applied to $S_0 = \P(\C_T)$ shows that the expected number of steps of~$\Epoch_2$ in which \ICB changes its partition is $O(\log |\P(\C_T)|)$.

    In the remaining steps of $\Epoch_2$, \ICB keeps its partition, which is then $\C_t$-preserving, and hence it pays nothing. Whenever \ICB changes its partition, it pays at most $n+1$ (cf.~\cref{lem:partition_preserving_trivial}), and thus
    \begin{align*}
        \E[\ICB(\Epoch_2)] 
        & = (n+1) \cdot O(\log |\P(\C_T)|) \\
        & = O(n \cdot (d \cdot \log (n/d) + q^2 \cdot \log q + n / q)),
    \end{align*}
    where the last equality follows by \cref{lem:P_C_T_bound}.
\end{proof}


\section{Final remarks}

In this paper, we provided the first algorithm for the online bisection problem with the competitive ratio of $o(n^2)$. 

The term $n^2 / q$ in our analysis stems from the number of large components, which can be as high as $n / q$: merges of two large components may be expensive. Improving this bound further seems to require an algorithm that handles components of medium sizes with the same care as \ICB handles components of sizes from the core set. Extending the result to a more general setting of online balanced graph partitioning (i.e., multiple-cluster case) is an intriguing open problem.

We note that our algorithm \ICB has non-polynomial running time; whether a~subquadratic competitive ratio is achievable in polynomial time without resource augmentation remains open.


\appendix


\section{Proof of \texorpdfstring{\cref{cla:number_theory}}{Claim \ref*{cla:number_theory}}}
\label{sec:proofs}

For a set $\set{s_1, \dots, s_m}$ of positive integers, we write 
\[
    \genmult{s_1,\dots,s_m} \define \left\{ 
        \textstyle
        \sum_{t=1}^m \lambda_t \cdot s_t \suchthat \lambda_t\in\Int_{\ge 0}
    \right\}.
\]
Thus $\genmult{s_1,\dots,s_m}$ is the set of all non-negative integer combinations of $s_1,\dots,s_m$, i.e., the numerical semigroup generated by $s_1,\dots,s_m$. The classical bound of Schur on the Frobenius number (see, e.g., \cite[Theorem 3.1.1]{Ramire05}) is given in the following claim.

\begin{claim}
\label{cla:generated_above_quadratic}
    Fix positive integers $s_1 < s_2 < \dots < s_m$ with $\gcd(s_1,\dots,s_m) = 1$.
    Let $S \define \genmult{s_1,\dots,s_m}$. The largest integer not in $S$ is at most 
    $(s_1-1) \cdot (s_m-1) < s_m^2$.
\end{claim}

\begin{proof}[Proof of \cref{cla:number_theory}]
    Let $g = \gcd(A \uplus B)$.

    It suffices to prove the claim in the case $g=1$. Indeed, if $g>1$, then we can divide all elements of $A$ and $B$, as well as $x$, by $g$. This does not change the coefficients, and the resulting equation will be $\sum_{i=1}^k \alpha_i \cdot (a_i/g) = x/g + \sum_{j=1}^\ell \beta_j \cdot (b_j/g)$, with both sides guaranteed to be at most $4 \cdot (H/g)^2$. By multiplying both sides by $g$, we obtain the original equation, with both sides guaranteed to be at most $4 \cdot (H/g)^2 \cdot g \leq 4 \cdot H^2$. Thus, in the following, we assume that $g = 1$.

    We start by defining 
    
    \begin{align*}
        g_A \define \gcd(A) 
        && S_A\define \genmult{a_1/g_A,\dots,a_k/g_A} 
        && u_A\define\max_i (a_i/g_A) \\
        g_B \define \gcd(B)
        && S_B\define \genmult{b_1/g_B,\dots,b_\ell /g_B}
        && u_B\define\max_j (b_j/g_B)
    \end{align*}
    Now let
    \[
        T \define \max \left\{
            u_A^2,
            \left\lceil (g_B \cdot u_B^2+x)/g_A \right\rceil
        \right\}.
    \]
    By our assumption, $g = 1$, and thus $\gcd(g_A,g_B) = 1$. Hence, the multiplication by $g_A$ permutes the residue classes modulo $g_B$. In particular, we have 
    \[
        \set{ (T + z) \cdot g_A \bmod g_B \suchthat z \in \{0, \dots, g_B-1 \} } = \set{0,\dots,g_B-1}.
    \]
    This means that there exists (exactly one) integer $t \in \set{T, \dots, T+g_B-1}$ 
    satisfying the congruence
    \[
        t \cdot g_A \equiv x \pmod{g_B}.
    \]
    Since $t\ge u_A^2$, \cref{cla:generated_above_quadratic} implies that $t\in S_A$. That is, $t = \sum_{i=1}^k \alpha_i \cdot (a_i/g_A)$ for some non-negative integers $\alpha_i$. Multiplying by $g_A$, we get
    \[
        \sum_{i=1}^k \alpha_i \cdot a_i = g_A \cdot t.
    \]
    As $t \cdot g_A \equiv x \pmod{g_B}$, the expression $(g_A \cdot t-x)/g_B$ is an integer. Furthermore, by the choice of $t$ and $T$, we have 
    $(g_A \cdot t-x)/g_B \geq (g_A \cdot T-x)/g_B \geq u_B^2$.
    Thus, \cref{cla:generated_above_quadratic} yields $(g_A \cdot t-x)/g_B\in S_B$, i.e., there exist non-negative integers $\beta_j$ such that
    \[
        \frac{g_A \cdot t-x}{g_B}=\sum_{j=1}^\ell \beta_j \cdot (b_j/g_B).
    \]
    That is,
    \[
    \sum_{i=1}^k \alpha_i \cdot a_i = g_A \cdot t = x+\sum_{j=1}^\ell \beta_j \cdot b_j.
    \]
    It remains to upper-bound the value of $g_A \cdot t$. Since $t\le T+g_B-1$, 
    \begin{align*}
        g_A \cdot t
        & \leq g_A \cdot T+g_A \cdot (g_B-1) \\*
        & = g_A \cdot \max \left\{
            u_A^2,
            \left\lceil (g_B \cdot u_B^2+x) / g_A \right\rceil
        \right\} + g_A \cdot (g_B-1) \\*
        & \leq \max \left\{
            g_A \cdot u_A^2,
            g_B \cdot u_B^2+x + g_A 
        \right\} + g_A \cdot g_B \\*
        & \leq \max \left\{
            H^2,\,
            H^2 + H + H 
        \right\} + H^2 \\*
        & \leq 4 \cdot H^2.
        && \qedhere
    \end{align*}
\end{proof}


\section{Proof of \texorpdfstring{\cref{cla:random_survivor}}{Claim \ref*{cla:random_survivor}}}
\label{sec:mts}

\begin{proof}[Proof of \cref{cla:random_survivor}]
    We first observe that if, at the beginning of round $j$, the algorithm's state is distributed uniformly over $S_{j-1}$, then two properties hold. First, the probability that it moves in round $j$ is $(|S_{j-1}| - |S_j|) / |S_{j-1}|$. Second, its state at the end of round $j$ is distributed uniformly over~$S_j$.
    
    The algorithm's initial state is a~single fixed state, and thus we simply charge its first move (if any) to the additive term~$1$. Let $j_0$ be the round of that move; by the argument above, from the end of round $j_0$ onwards, the algorithm's state is uniform over the current set of alive states. Hence, the expected number of subsequent moves is at most
    \[
        \sum_{j > j_0} \frac{|S_{j-1}| - |S_j|}{|S_{j-1}|}
        \leq \sum_{j > j_0} \; \sum_{i = |S_j|+1}^{|S_{j-1}|} \frac{1}{i}
        \leq \sum_{i=2}^{|S_0|} \frac{1}{i}
        \leq \ln |S_0|,
    \]
    where the first inequality holds as each of the $|S_{j-1}| - |S_j|$ summands is at least $1 / |S_{j-1}|$, and the second one as the sets $S_j$ are nested.
\end{proof}



\bibliography{references}

\end{document}